\documentclass[11pt,prl,singleside,a4paper]{revtex4-2}

\usepackage{multirow}
\usepackage{xcolor}
\usepackage{bm}
\usepackage{amsfonts,amssymb,amsmath}
\usepackage{graphicx,dcolumn,bm,color,braket,slashed}
\usepackage{times} 
\usepackage{array}
\usepackage{hyperref}
\usepackage{textcomp}
\usepackage{siunitx}
\usepackage[normalem]{ulem}
\usepackage{todonotes}
\usepackage{cancel}
\usepackage[authormarkup=none]{changes}
\definechangesauthor[color=cyan]{SJ}

\usepackage{kotex}

\begin{document}

\title{Ultrafast Current Switching from Quantum Geometry in Semimetals}

\author{Youngjae Kim$^{\ast}$}
\affiliation{School of Physics, KIAS, Seoul 02455, Korea}
\affiliation{Department of Semiconductor Physics, Kangwon National University, Chuncheon 24341, Republic of Korea}

\author{Sejoong Kim$^{\ast}$}
\affiliation{Department of Electronic and Electrical Convergence Engineering, Hongik University, Sejong 30016, Republic of Korea}

\author{Jun-Won Rhim$^{\ast}$}
\affiliation{Department of Physics, Ajou University, Suwon 16499, Korea}

\email{Corresponding authors}
\email{ykim.email@gmail.com}
\email{sejoong@hongik.ac.kr}
\email{jwrhim@ajou.ac.kr}

\begin{abstract}

Technological progress towards next-generation electronics critically relies on achieving faster switching with reduced energy consumption. Because device operation speeds are fundamentally constrained by the intrinsic properties of constituent materials, identifying systems with inherently superior switching capabilities is essential.
Here, we propose that semimetallic systems characterized by non-trivial quantum geometry, including quadratic band-touching semimetals and singular flat bands, can serve as a promising platform for ultrafast switching at voltages compatible with modern electronics.
We show that, in such quantum geometric semimetals, an electric current is generated instantaneously upon application of a moderate external electric field, reaching its steady-state value.
As a consequence, the current exhibits rapid and stable on–off switching behaviour under periodic optical pulse trains, demonstrating robustness under experimentally feasible conditions.
In terms of switching speed, this quantum geometric semimetal outperforms conventional metals, semiconductors, and graphene.
We identify the microscopic origin of this behaviour as interband coupling governed by the Hilbert–Schmidt quantum distance, together with a finite density of states at the band-touching point. This mechanism further leads to a universal classification of conductivity for both gapless and gapped quantum geometric semimetals.
Finally, first-principles calculations suggest realistic material platforms, including bilayer graphene, cyclic graphene, monolayer bismuth and V$_3$F$_8$—in which the predicted instantaneous current switching can be directly realized, further supported by time-dependent density functional theory simulations performed for representative systems.
\end{abstract}

\maketitle

\section{Introduction}
The switching frequency of electric current determines signal processing speed in electronics\cite{ProcRSocLond.145.523-529,PhysRevLett.59.1018, Nanometer.CMOS.2010}. 
Over the past decade, the increasing demand for faster signal processing has posed significant challenges in electronics, quantum computing, and information technology~\cite{LED.2015.2407193,Nat.Phys.12.724-725.2016,Nat.Rev.Phys.Heide2024,Yoshioka2024natele}. 
While recent technical advancements continue to push the boundaries of performance, the operating frequency of these electronics is fundamentally constrained to a limited frequency range, mainly due to carrier relaxation processes~\cite{SSE.51.1079-1091.2007, Nat.Phys.12.724-725.2016}.
Relaxation processes introduce unavoidable dissipation during electric current switching, primarily induced by phonons~\cite{Elect.Eng.in.Jap.211.40-46.2020,NewJPhys21.093005,Phys.Rev.B97.205412}. 
Therefore, this dissipation manifests as a switching time delay of 0.1 to 1 ps (1 ps $=10^{-12}$~s)~\cite{Solid.State.Electronics.44.35-340.2000,Sci.Rep.4.4886.2014,Nat.Photon.8.205-213.2014, J.Appl.Phys.58.857-861.1985,Elect.Eng.in.Jap.211.40-46.2020,NewJPhys21.093005,Phys.Rev.B97.205412}, which fundamentally limits the operating frequency below the terahertz (THz, $10^{12}$ Hz) range~\cite{Nat.Commun.6.7655.2015,New.J.Phys.23.063047.2021,Elect.Eng.in.Jap.211.40-46.2020,PhysRevB.84.125450,PhysRevB.97.205412,Nat.Phys.12.724-725.2016, Nat.Rev.Phys.Heide2024}.

%
Recently, all-optical light-wave control has emerged as a promising 
alternative, capable of reaching terahertz-to-petahertz operating 
speeds~\cite{Nat.Rev.Phys.Heide2024, Nature.493.70-74.2013, Nat.Phys.12.724-725.2016, 
Nature.550.224-228.2017,Nat.Commun.13.1620.2022,Light-Shaping,Dixit1,Dixit2}. 
In this regime, an optical pulse directly drives coherent electronic motion, 
yet fundamentally requires extremely strong electric fields $\sim$ V/nm 
($\sim10^5$ kV/cm)~\cite{nature_rev_matt} to accelerate electrons within the bands. 
As a result, such approaches rely on intense fields~\cite{Nat.Rev.Phys.Heide2024}, posing a significant barrier to their implementation in conventional electronic platforms, which typically operate in the $\sim$ kV/cm field regime~\cite{nature_rev_matt}.
Therefore, an approach that achieves ultrafast switching at substantially 
weaker fields remains highly desirable for integration into 
modern electronic platforms.
Given these considerations, it would be timely to suggest an emerging class of systems that opens new pathways for a comprehensive restructuring of electric current generation.

In this study, we propose that quantum-geometrically characterized semimetals could provide an advanced platform for this pursuit.
We call such a semimetal a quantum geometric semimetal (QGS).
Here, quantum geometry refers to the geometric structure of the Hilbert space formed by the Bloch eigenstates, where distances between states are measured using the Hilbert-Schmidt quantum distance~\cite{PhysRevA.54.1844,Dodonov01032000}. 
This distance is defined as
\begin{align}
d^2_\mathrm{HS} = 1 - |\langle v_\mathbf{k} | v_{\mathbf{k}^\prime} \rangle|^2,\label{eq:quantum_distance}
\end{align}
where $v_\mathbf{k}$ is the eigenvector of the Bloch Hamiltonian at crystal momentum $\mathbf{k}$.
The QGSs are distinguished from conventional topological semimetals in the following way.
In two-dimensional Dirac semimetals, the Dirac node is topologically characterized by a quantized Berry phase of $\pi$, while in Weyl semimetals, the Weyl point carries a quantized monopole charge.
In contrast, the quadratic band-touching (QBT) point in a singular flat band (SFB) system—one of the QGSs—is characterized geometrically by the maximum quantum distance ($d_\mathrm{max}$) near the touching point~\cite{Nature.584.59-63}.
More precisely, $d_\mathrm{max}$ denotes the maximum value of $d^2_\mathrm{HS}$ among all pairs of Bloch eigenstates with momenta $\mathbf{k}$ and $\mathbf{k}^\prime$ near the band-crossing point (BCP).
At the BCP, the Bloch eigenstate exhibits a discontinuity, which underlies the designation of the system as a singular flat band~\cite{PhysRevB.99.045107,rhim2021singular,oh2022bulk}.
Likewise, the QBT point in 2D QGSs (Figs.~\ref{FIG1}\textbf{a} and \textbf{b}) is also characterized by $d_\mathrm{max}$ or by the elliptic parameters of the projected elliptic image formed by the Bloch wave functions in the vicinity of the band-touching point~\cite{PhysRevB.103.L241102,oh2025universal}.
Importantly, unlike topological invariants, these geometric quantities vary continuously.
Quantum geometry has emerged as a cutting-edge concept in condensed matter physics, innately linked to the Berry curvature and the quantum metric \cite{PhysRevB.81.245129,yu2024quantum}. 
While the Berry curvature has been studied exhaustively over the past decades in the context of quantum Hall and valley Hall effects \cite{PhysRevLett112.166601,Science2023.381.6654,Science2014.344.6191,Science2022.375.6587}, the quantum metric and distance have recently gained attention for their role in unveiling various physical properties. 
For instance, the quantum distance has been shown to play a key role in the Landau level spacing of bands with the QBT~\cite{Nature.584.59-63,oh2024revisiting}, interface states of singular flat bands \cite{CommunicationsPhysics5.320}, thermoelectric transport \cite{AdvSci2024.11.2411313}, and optical properties \cite{PhysRevLett129.227401,SciAdv10.51,NatCommun8.14176}. 
On the other hand, the nontrivial quantum metric is related to the superfluid weight~\cite{torma2022superconductivity,peotta2015superfluidity,liang2017band,verma2024geometric,HuPRB2022}, current noise~\cite{neupert2013measuring}, exciton properties~\cite{jankowski2024excitonic}, polarizability~\cite{Resta2006,verma2024quantum,bouhon2303quantum}, and capacitance~\cite{komissarov2024quantum}.

Here, we show that quantum geometry intrinsically enables superior electric current switching.
From the time-dependent current response to a step-like electric field (Fig.~\ref{FIG1}\textbf{c}), the rise time in QGSs significantly outperforms that of conventional metals, semiconductors, and graphene.
The rise time, defined as the time-interval for the current to increase from 10\% to 90\% of its steady-state value, quantifies the switching speed~\cite{Semiconductor.Physics.and.Devices}. 
This nearly instantaneous response originates from Schwinger-like pair creation~\cite{PhysRevB81.165431}, enabled by interband coupling governed by the quantum distance and the finite density of states at the QBT.
In the steady state, the conductivity follows a universal form, $\sigma = e^2 d_{\rm{max}}^2 / 8\hbar$, independent of material-specific parameters such as effective masses or band-touching details.
Importantly, this ultrafast response persists under time-dependent driving: under optical pulses, QGSs exhibit equally rapid switching, remaining robust up to petahertz frequencies with required field strengths of a few kV/cm.
Density functional theory calculations further confirm robust switching behavior in realistic QGS materials.
These results establish QGSs as promising platforms for ultrafast current switching and highlight the role of quantum geometry in nonequilibrium transport.

\section{Results}

\subsection{Quantum Geometric Semimetals and Real-Time Dynamics}
To demonstrate microscopic current generation and switching capabilities, we simulate the real-time dynamics of electrons in QGSs under an abruptly applied electric field, and extend the analysis to more general optical driving fields for continuous switching under realistic conditions. 
The dynamics are described by a quantum master equation incorporating relaxation and dephasing processes (see Appendix for details). 
We consider a generic continuum model for quantum geometric semimetals (QGSs) characterized by $d_\mathrm{max}$ and isotropic band dispersions (see Appendix).

\begin{figure*}[t]
\includegraphics[width=1.0\textwidth]{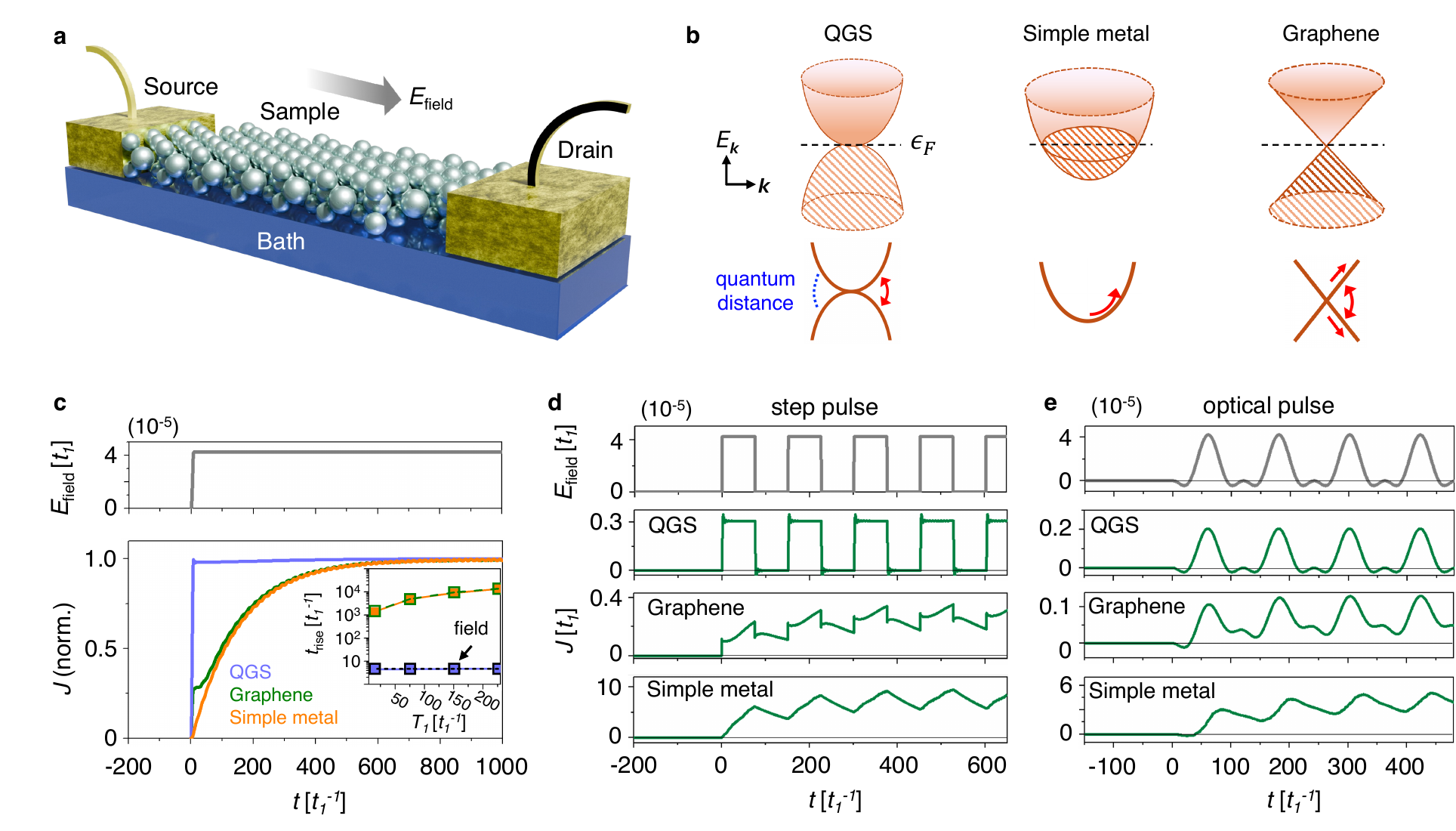}
\caption{
\textbf{Real-time current generations and electric current switching.}
(a) Schematic illustration of the current measurement in a two-terminal setup. An external electric field $E_{\mathrm{field}}$ is applied between the terminals through the sample, which is coupled to the bath.
(b) 
We consider three representative cases for the sample region: the quantum geometric semimetal (QGS), a simple metal (isolated quadratic band), and graphene (linear Dirac bands).
Hatched areas below the Fermi energy $\epsilon_{F}$ represent initially occupied bands, and red arrows denote electron pathways contributing to the current. 
(c) (Upper panel) The time profile of $E_{\mathrm{field}}$ applied to the sample. 
(Lower panel) Normalized time-dependent current densities for QGS (blue), simple metal (orange), and graphene (green). The inset compares the rise times for the QGS, simple metal, and graphene. The rise time is defined as $t_{\mathrm{rise}} = t_{90\%}-t_{10\%}$, where $t_{X\%}$ is the time to reach $X\%$ of the final steady state current. 
(d,e) Current switching response of the three systems to (d) a sequence of square pulses and (e) a sequence of optical pulses of $E_{\mathrm{field}}$.
The bath coupling parameters are set to $T_{1} = 150/t_1$ and  $T_{2} = 30/t_1$, unless otherwise stated.
}
\label{FIG1}
\end{figure*}

We apply the continuum Hamiltonian of the QGS to the master equation to analyze the time-dependent current generation under an abruptly applied electric field (see Appendix).
Once we obtain the density matrix $\rho_\mathbf{k}(t)$, the electric current is computed by $\mathbf{J}(t) = (1/2\pi)^2\int d^{2}\mathbf{k}\left\langle {\mathbf{j}}_{\mathbf{k} - \frac{e}{\hbar c} \mathbf{A}(t)} \right\rangle$, where $\left\langle {\mathbf{j}}_{\mathbf{k} - \frac{e}{\hbar c} \mathbf{A}(t)} \right\rangle = \mathrm{Tr}[\rho_{\mathbf{k}}(t){\mathbf{j}}_{\mathbf{k} - \frac{e}{\hbar c}\mathbf{A}(t)}]$ with the current operator ${\mathbf{j}}_{\mathbf{k} - \frac{e}{\hbar c}\mathbf{A}(t)} = -e\partial_{\mathbf{k}}H_{\mathbf{k}-\frac{e}{\hbar c}\mathbf{A}(t)}$.
Expanding the current density to linear order in the electric field, we obtain
\begin{align}
\mathbf{J}_\mathrm{QGS}(t) = \frac{e^2}{\hbar}\frac{d^2_\mathrm{max}}{8}E_\mathrm{field} \Theta(t),\label{eq:step_current}
\end{align}
where the fundamental constants $e$ and $\hbar$ are reinstated.
Here, the Fermi energy $\epsilon_{F}$ is set to the band-crossing point (BCP) so that the lower band is initially occupied. 
This analytical expression accurately captures the current dynamics obtained from numerical simulations, as shown in the lower panel of Fig.~\ref{FIG1}\textbf{c}.
The time-dependent electric currents are numerically computed using the quantum master equation within the Houston basis~\cite{PhysRevLett.73.902,1361-6455/abb127}.

Two points are worth noting.
First, in response to the step-like electric field, the current immediately reaches its steady-state value ($E_\mathrm{field}e^2 d^2_\mathrm{max}/8\hbar$) without delay.
Namely, the rise time is zero~\footnote{In numerical calculations, a nearly step-like electric field with a finite switching duration is used, leading to a small but finite rise time. As the switching duration is reduced, the rise time approaches zero.}.
This behavior contrasts sharply with that of conventional simple metals and graphene.
A simple metal is modeled by a quadratic band of a doped semiconductor, while graphene exhibits linear bands crossing at the Dirac point (see Appendix).
Applying the same framework to these systems yields time-dependent current profiles that gradually approach steady states, with finite rise times governed by $T_1$, as shown in Fig.~\ref{FIG1}\textbf{c}.
Second, finite current generation in QGSs requires a nonzero $d_\mathrm{max}$, indicating that transport is governed by quantum geometry.
A more detailed analysis is presented in the next subsection.

We demonstrate that the instantaneous current generation in QGSs originates from a fundamentally different mechanism: the current is predominantly of interband origin, while the intraband contribution is strongly suppressed, as illustrated in Fig.~\ref{FIG1}\textbf{b}.
In contrast, electric current in a simple metal arises from intraband processes, where a population imbalance within the conduction band generates a current governed by the time scale $T_1$.
The interband and intraband currents are computed by the decomposition of current into $\mathbf{J}^\mathrm{ter}(t) = (1/2\pi)^2\int d^{2}\mathbf{k}\mathrm{Tr}[\rho^{\mathrm{ter}}_{\mathbf{k}}(t){\mathbf{j}}_{\mathbf{k} - \frac{e}{\hbar c}\mathbf{A}(t)}]$ and $\mathbf{J}^\mathrm{tra}(t) = (1/2\pi)^2\int d^{2}\mathbf{k}\mathrm{Tr}[\rho^{\mathrm{tra}}_{\mathbf{k}}(t){\mathbf{j}}_{\mathbf{k} - \frac{e}{\hbar c}\mathbf{A}(t)}]$, respectively,
where $\rho^{\mathrm{ter}}_{nm,\mathbf{k}}(t)=(1-\delta_{nm})\rho_{nm,\mathbf{k}}(t)$ and $\rho^{\mathrm{tra}}_{nm,\mathbf{k}}(t)=\delta_{nm}\rho_{nm,\mathbf{k}}(t)$.
In QGSs, the density of states remains finite near the BCP, enabling strong interband current generation via Schwinger-like electron-hole pair creation~\cite{PhysRevB81.165431,PhysStatusSolidiB248.11}.
Since this process does not require any time delay, current generation occurs immediately upon application of the electric field (see Supplementary Sec.~4).
In contrast, the intraband current is suppressed due to the nearly vanishing group velocity around the BCP.
These features render QGSs particularly suited for switching applications, exhibiting an essentially instantaneous response with infinitesimal $t_{\rm{rise}}$ (see Supplementary Sec.~6A).
In simple metals, interband processes are absent, and the current arises solely from intraband contributions of free carriers with density $n_e$, ${J}(t) = 4t_1 n_e T_1 (1 - e^{-t/T_1}) E_\mathrm{field}$, requiring a classical acceleration time set by $T_1$ and resulting in a finite rise time $t_{\rm{rise}}\sim T_1$.
In graphene, although interband processes are allowed, the vanishing density of states at the BCP suppresses them, while the large group velocity enhances intraband contributions, resulting in a mixed response.
Consequently, graphene exhibits an initial partial current jump followed by a gradual increase governed by $T_1$ (see Supplementary Sec. 7 and Sec. 8), as shown in Fig.~\ref{FIG1}\textbf{c}.
In contrast, QGSs exhibit instantaneous responses independent of $T_1$, highlighting a fundamentally distinct mechanism of current generation.

The ultra-short rise time of QGSs implies extremely fast turn-on/off responses of electric current.
To demonstrate this, we compute current profiles under a sequence of step pulses, as shown in Fig.~\ref{FIG1}\textbf{d}.   
The current in QGSs immediately follows the applied field, forming a square-wave response with sharply defined on/off regions synchronized with the field.
Thus, QGSs enable ultrahigh-frequency switching controlled solely by the electric field. 
In contrast, simple metals and graphene exhibit finite rise and fall times, causing delayed responses, residual current during turn-off, and poorly defined switching regions, as shown in Fig.~\ref{FIG1}\textbf{d}.
These behaviors originate from the dominance of intraband current generation in both systems.

Beyond step pulses, we consider optical pulses with oscillating electric fields to examine experimental feasibility, as shown in Fig.~\ref{FIG1}\textbf{e}.
Similarly, QGSs exhibit current responses that closely follow the applied field, with clearly separated on/off regions, whereas the other systems do not.
Notably, this switching occurs at field strengths approximately $10^5$ times lower than those required for all-optical light-wave current generation~\cite{nature_rev_matt}.

\begin{figure*}[t]
\includegraphics[width=1.0\textwidth]{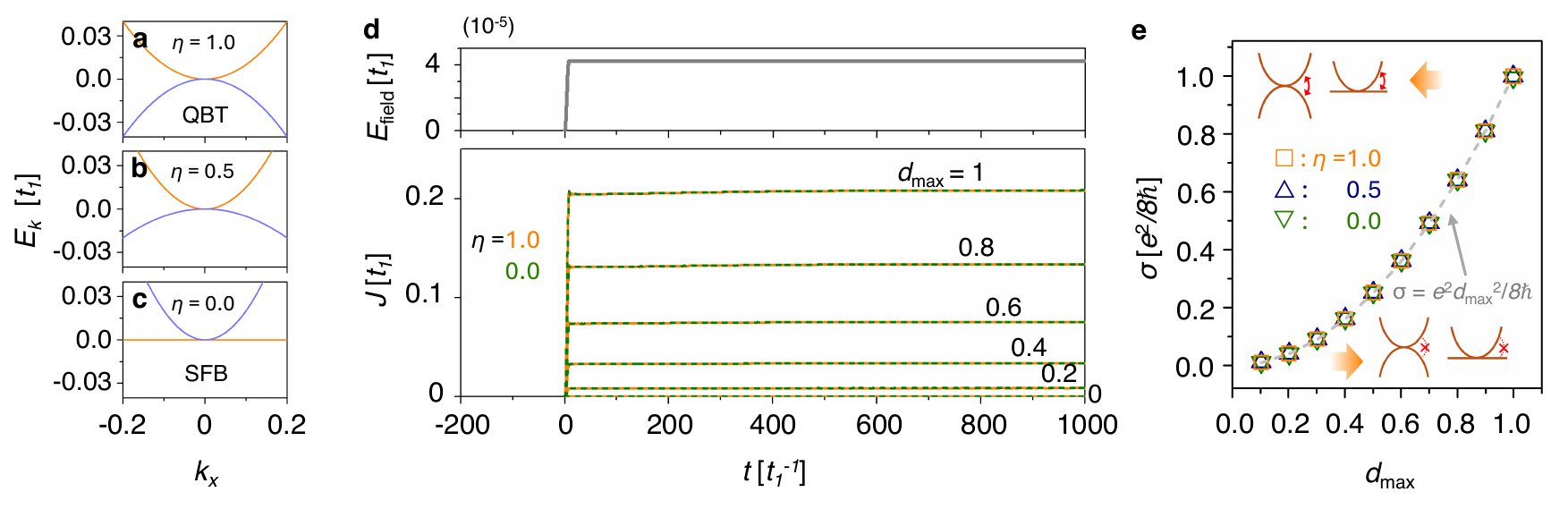}
\caption{
\textbf{Universal DC conductivity of quantum geometric semimetals.}  
(a–c) Electronic structures of quadratic band touching systems for different values of $\eta$.  
Panels (a), (b), and (c) correspond to $\eta = 1.0$, $0.5$, and $0$, representing symmetric touching (QBT), asymmetric touching, and singular flat band (SFB), respectively. 
(d) Time-dependent current profiles for systems with $\eta = 1$ and $\eta = 0$, computed for various values of $d_\mathrm{max}$.  
(e) Relationship between the steady state conductivity, $\sigma_0 = J(t\rightarrow\infty)/E_{\mathrm{field}}$, and $d_\mathrm{max}$.
The gray line represents the analytic curve $\sigma = e^2d_\mathrm{max}^{2}/8\hbar$, while orange boxes, blue upper triangles, and green lower triangles denote numerical results for $\eta = 1$, $\eta = 0.5$, and $\eta = 0$, respectively. The bath coupling parameters are set to $T_{1} = 150/t_1$ and $T_{2} = 30/t_1$.}
\label{FIG2}
\end{figure*}

\subsection{Universal Steady State Conductivity}

In this subsection, we examine the quantum geometric origin of the DC current in QGSs by analyzing the role of $d_\mathrm{max}$.
To this end, we extend the discussion to a broader parameter range of $d_\mathrm{max}$ and $\eta$.
Here, $\eta$ controls the ratio of the effective masses of the conduction and valence bands, as shown in Fig.~\ref{FIG2}\textbf{a}–\textbf{c}: symmetric QBT ($\eta = 1$), asymmetric touching ($\eta = 0.5$), and SFB ($\eta = 0$) (see Appendix).
The real-time DC current responses for various $d_\mathrm{max}$ and $\eta$ are presented in Fig.~\ref{FIG2}\textbf{d}.

First, for fixed $d_\mathrm{max}$, the current response is independent of $\eta$.
Although only two representative cases ($\eta = 1$ and $0$) are shown, identical behavior is obtained across the full range of $\eta$.
This is because $\eta$ enters the Hamiltonian only as a prefactor to the identity matrix and therefore does not affect the dynamics when the Fermi level lies at the BCP.
Thus, as long as the bands remain quadratically touching, the resulting DC current and conductivity are independent of $\eta$.

In contrast, the quantum distance $d_\mathrm{max}$ plays a central role in determining the DC current, as shown in Eq.~(\ref{eq:step_current}).
The DC current scales as $d_\mathrm{max}^2$, reflecting the strength of interband coupling responsible for Schwinger-like electron-hole pair generation~\cite{PhysRevB81.165431, TANJI20091691}.
This interband (polarization) current originates from the dipole matrix element, which is also proportional to $d_\mathrm{max}^2$ (see Supplementary Sec.~2).
This geometric behavior is further confirmed by the steady-state conductivity $\sigma_0 = J(t \rightarrow \infty)/E_\mathrm{field}$ shown in Fig.~\ref{FIG2}\textbf{e}.
For all values of $\eta = 1.0, 0.5,$ and $0$, the conductivity curves coincide and exhibit quadratic scaling with $d_\mathrm{max}$.
The analytical expression $\sigma = e^2 d_\mathrm{max}^2 / 8\hbar$ (see Supplementary Sec.~5) agrees perfectly with the numerical results, demonstrating universal DC conductivity.
These results show that DC transport in QGSs is governed by intrinsic quantum geometry, independent of $\eta$.

\begin{figure*}[ht]
\includegraphics[width=1.0\textwidth]{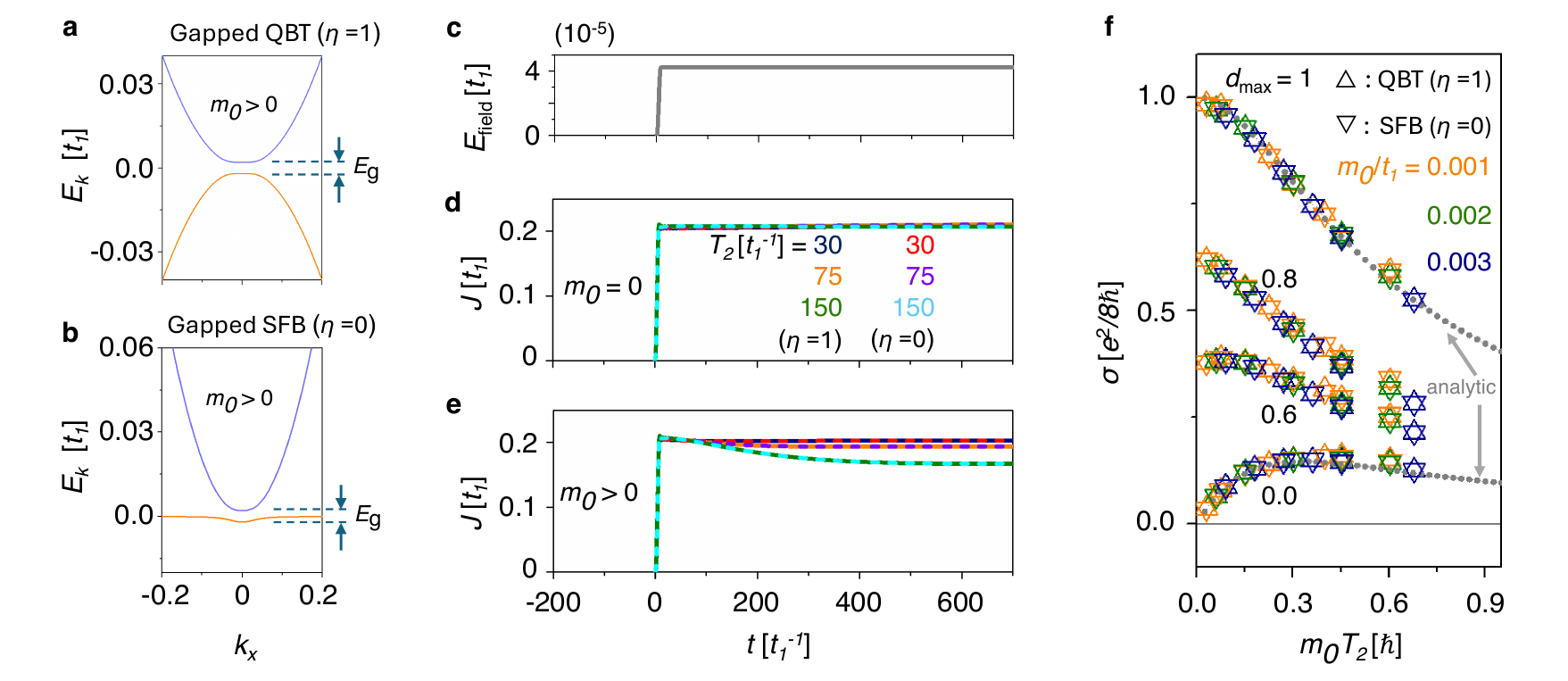}
\caption{
\textbf{Conductivity characteristics of gapped QBT systems.}  
(a, b) Electronic structures of gapped systems for (a) QBT ($\eta=1$) and (b) SFB ($\eta=0$), with $m_0/t_1 = 0.002$ and a band gap of $E_g = 2m_0$.  
(c) Time profile of the DC bias electric field. 
(d,e) Time-dependent current profiles for (d) gapless ($m_0=0$) and (e) gapped ($m_0>0$) systems with $d_\mathrm{max} = 1$ when $T_1 = 150 / t_1$, and $T_2=30/t_1, 75/t_1, 150/t_1$. QBT ($\eta = 1$, solid lines) and SFB ($\eta = 0$, dashed lines) are compared in the same panel.  
(f) Conductivity characteristic curves as a function of $m_0 T_2$, varying  $d_{\text{max}}$ and $m_0$.  
Upper and lower triangles are full numerical results of SFB and QBT, respectively. The triangle color is associated with the value of the mass term $m_0$. 
Two gray dotted lines represent analytic conductivity curves for $d_\mathrm{max} = 1$ (upper) and $d_\mathrm{max} = 0$ (lower). 
The conductivity on $d_\mathrm{max}$ and $m_0T_2$ by explicitly showing that it takes the form $\sigma = \frac{e^2}{8\pi^{2}T_{2}\hbar}\int d\mathbf{k}^2 \mathcal{G}(\mathbf{k}, d_\mathrm{max}, m_0T_2)$, where the integrand $\mathcal{G}(\mathbf{k}, d_\mathrm{max}, m_0T_2)$ depends solely on $d_\mathrm{max}$ and $m_0T_2$ and is defined in the Appendix section.
For $d_\mathrm{max} = 1$, its leading order is given by $\sigma=\frac{e^2}{2\pi\hbar}m_0T_2\left(1-2m_0T_2\text{acot}(2m_0T_2)\right)$, and for $d_\mathrm{max} = 0$, by $\sigma=\frac{e^2}{8\pi\hbar}[2(1-4m_0^2T_2^2)\text{acot}(2m_0T_2)+4m_0T_2]$.
See Supplementary Sec.~5 for detailed derivations.
}
\label{FIG3}
\end{figure*}

\subsection{Gapped systems}

So far, we have examined the robustness of instantaneous current generation and the corresponding conductivities in gapless QGS systems.
In real materials, however, the BCP can be slightly gapped due to spin–orbit interaction or broken inversion symmetry.
To unify our analysis, we extend the discussion to include gapped systems.
As in the gapless case, the conductivity remains independent of the effective masses of the conduction and valence bands and of $T_1$.
However, once a gap opens, it additionally depends on the gap size and $T_2$, along with $d_\mathrm{max}$.
We find that the conductivity of gapped QGSs is fully determined by $d_\mathrm{max}$ and the product of the band gap and $T_2$.

Fig.~\ref{FIG3}\textbf{a} and \textbf{b} show the electronic structures of QBT and SFB systems, respectively, for a small $m_0$, corresponding to a slight band gap $E_g = 2m_0$.  
For comparison, under a step-like electric field (Fig.~\ref{FIG3}\textbf{c}), we plot the time-dependent current profiles of gapless QBT and SFB systems for various dephasing times $T_2$ in Fig.~\ref{FIG3}\textbf{d}. 
The current response in gapless systems is independent of $T_2$.
Since the population decay time $T_{1}$ also has no effect (Fig.~\ref{FIG1}\textbf{b}), the current is robust against system–bath coupling for gapless cases.
In contrast, in gapped systems, the current gradually decreases from its initial value and saturates below $e^2 d_\mathrm{max}^2 /8\hbar$, as shown in Fig.~\ref{FIG3}\textbf{e}.
This behavior originates from the absence of density of states at the BCP (see Supplementary Sec.~4C) and becomes more pronounced with increasing $T_2$.

While the time evolution of the current differs between gapless and gapped systems (Figs.~\ref{FIG3}\textbf{d}~and~\textbf{e}), both exhibit an infinitesimal rise time, with the current immediately reaching its maximum upon application of a sudden electric field.
In conventional semiconductors, where the current monotonically approaches saturation via intraband processes, the rise time directly quantifies the time to reach the steady-state current. 
In contrast, in gapped QBT systems, the rise time does not reflect the time to reach the steady state; instead, the current shows a nonmonotonic evolution, relaxing to the steady state over a much longer timescale. 
Nevertheless, the initial current surge, which transiently exceeds the steady-state value, can still define the switching-on state.

In Fig.~\ref{FIG3}\textbf{f}, we show that the conductivity of gapped QGSs exhibits a universal dependence on $d_\mathrm{max}$ and the product $m_0 T_2$.
By varying model parameters ($\eta$, $t_1$, $m_0$, $d_\mathrm{max}$) and system–bath couplings ($T_1$, $T_2$), we find that the conductivity is determined solely by $d_\mathrm{max}$ and $m_0 T_2$.
%
%
When a small band gap is introduced, QGSs can exhibit finite conductivity even at $d_{\mathrm{max}}=0$, since the gap-opening term provides an alternative source of interband coupling in the presence of finite dephasing.
However, in the limit $m_0 T_2 \gg \hbar$, that is, when the gap is sufficiently large or dephasing is nearly absent, the conductivity eventually decreases.

Since interband current is the only available transport channel when the Fermi level lies in the gap, the conductivity is independent of $T_1$.

\begin{figure*}[t]
\includegraphics[width=1.0\textwidth]{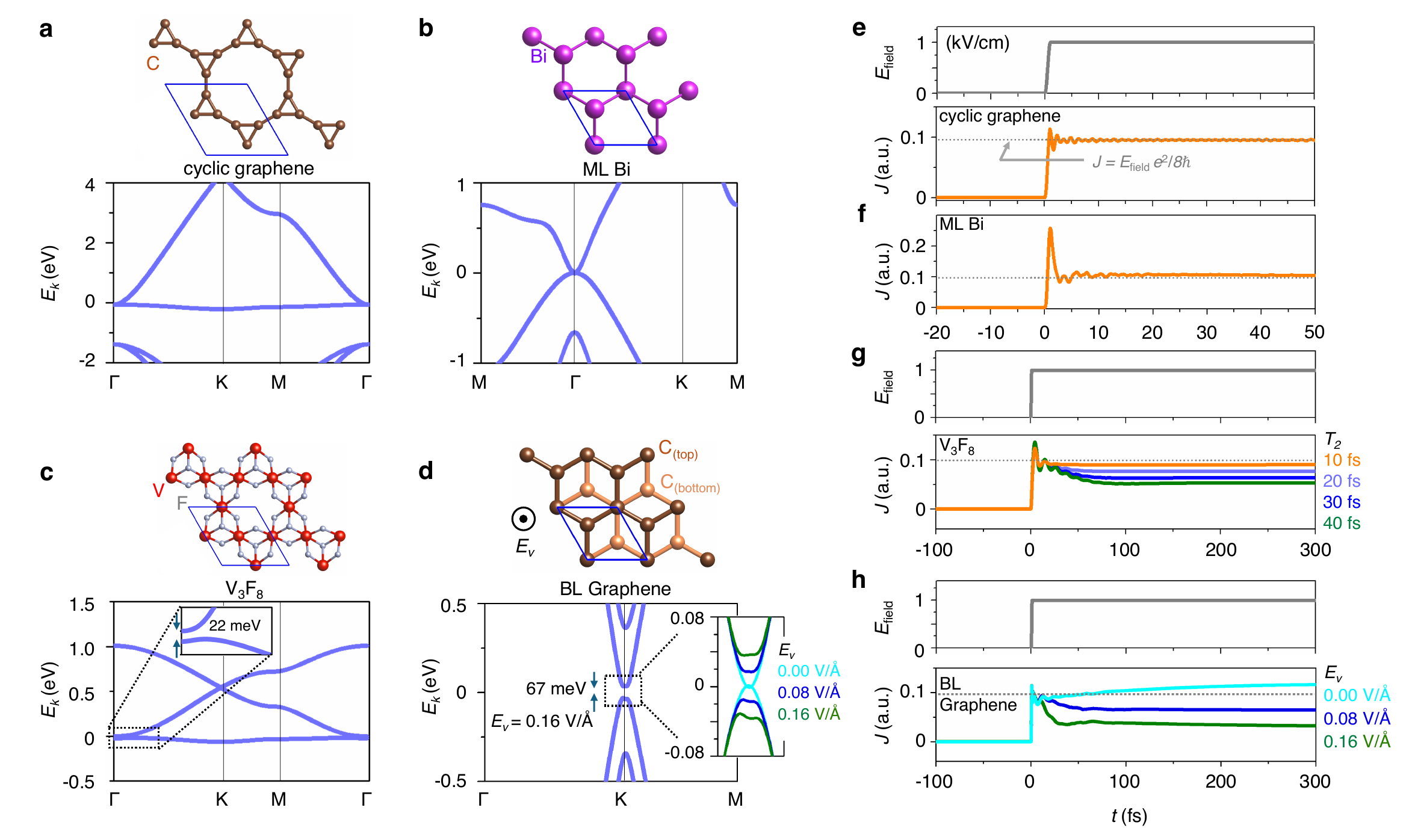}
\caption{
\textbf{Current responses in candidate materials.}
Crystal structures and electronic structures of (a) cyclic graphene, (b) ML Bi, (c) V$_{3}$F$_{8}$, and (d) BL graphene with AB stacking.
The Fermi level ($\epsilon_{F} = 0$ eV) is aligned with the valence band edge.
(e–h) Time-dependent current densities.  calculated under an applied bias electric field of 1 kV/cm for (e) cyclic graphene, (f) ML Bi, (g) V$_{3}$F$_{8}$, and (h) BL graphene.
Unless otherwise stated, the population decay time $T_1$ is fixed at 100 fs, and the dephasing time $T_2$ is set to 20 fs.
}
\label{FIG4}
\end{figure*}

\subsection{Candidate Materials: First-Principles Approaches}

We apply the general theory of instantaneous current generation in QGSs to realistic materials using first-principles calculations.
We consider four candidate two-dimensional systems: (a) cyclic graphene, (b) monolayer bismuth (ML Bi), (c) V$_3$F$_8$, and (d) bilayer (BL) graphene with AB (Bernal) stacking. 
Their lattice structures and band dispersions are shown in Figs.~\ref{FIG4}\textbf{a}–\textbf{d}. 
Cyclic graphene and V$_3$F$_8$ realize gapless and gapped SFB systems, respectively, both featuring kagome-like electronic structures arising from hexagonal lattices of triangular units.
This structure yields highly flat valence bands, effectively realizing the SFB scenario~\cite{Nature.584.59-63,PhysRevB.98.035135,acs.jpcc.3c07068}. 
In V$_3$F$_8$, spin–orbit coupling opens a small band gap of 22 meV.
In contrast, ML Bi and BL graphene realize the QBT model.
ML Bi has a finite gap at $\Gamma$, which can be closed by biaxial strain, yielding a gapless QBT system (Fig.~\ref{FIG4}\textbf{b}). 
BL graphene corresponds to a gapped QBT with a tunable gap controlled by an external vertical electric field $E_\mathrm{v}$. 
All four materials are QGSs with $d_{\text{max}}=1$.

To explore current generation, we perform time-evolution simulations based on the quantum master equation combined with first-principles calculations (see Appendix). 
We first examine two gapless systems: cyclic graphene and ML Bi.
An external electric field of 1 kV/cm is applied, and the resulting currents are shown in Figs.~\ref{FIG4}\textbf{e} and \textbf{f}.
This field range ($\sim$kV/cm) is typical in electronic devices~\cite{s10854-012-0818-2,T-ED.1972.17468, T-ED.1983.21088, 1.1999025} and is $10^4$–$10^5$ times smaller than that required for light-wave currents ($\sim$ $10^4$–$10^5$ kV/cm)~\cite{nature_rev_matt}.
In both systems, the current is generated almost instantaneously, consistent with the continuum predictions.
However, it exhibits damped oscillations before stabilizing within $\pm 10$\% of the saturated value, $E_\mathrm{field}e^2 d_\mathrm{max}^2 /8\hbar$ per spin and valley. 
As a result, both materials show a finite rise time of $\sim$2–3 fs ($1\,\mathrm{fs}=10^{-15}\,\mathrm{s}$).
This deviation from the ideal continuum behavior originates from the finite momentum cutoff $k_c$ in DFT band structures, imposed by the Brillouin zone and deviations from perfect parabolic dispersion, whereas the continuum model assumes $k_c \to \infty$ (see Supplementary Sec.~6A).
The rise-time expression is provided in Appendix.
Including spin and valley degeneracies, the above value corresponds to a total current density of $\sim$10–100 A/m under fields of 1–10 kV/cm, comparable to experimental values in 2D field-effect transistors~\cite{real_current_1,real_current_2}.

Next, we examine current generation in gapped QGSs, focusing on V$_3$F$_8$ and BL graphene under a perpendicular electric field, as shown in Fig.~\ref{FIG4}\textbf{g} and \textbf{h}.
We first consider V$_3$F$_8$, a gapped SFB system.
The simulations use dephasing times $T_{2}$ in the range of 10–40 fs, typical for two-dimensional materials~\cite{PhysRevB.103.L041408,science.aam8861,acs.nanolett.1c02538,nn504760x,New.J.Phys.23.063047.2021,NewJPhys21.093005,PhysRevB.84.205406,PhysRevA.106.033123}.
The steady-state current shows a clear dependence on $T_2$, with increasing resistivity as $T_2$ increases, deviating from the geometric current $J = E_\mathrm{field}e^2 d_\mathrm{max}^2 /8\hbar$. 
This behavior is consistent with that of gapped geometric semimetals discussed in Fig.~\ref{FIG3}.
Although V$_3$F$_8$ requires tens of femtoseconds to reach the steady state, with longer times for larger $T_2$, the rise time, set by the initial current jump, remains on the order of a few femtoseconds. 
This short rise time indicates its potential for ultrafast current generation.

As another example of a gapped QGS, we examine BL graphene.
This is a 2D QBT semimetal within the QGS framework of Eq.~(\ref{Eq1}), although this description strictly holds only for the simplest parabolic low-energy model ($H^q_{\mathbf{k}} \sim k^2$).
A more accurate description requires a small linear correction ($H^w_{\mathbf{k}} \sim \mathbf{k}$), leading to trigonal warping and linear dispersion near the BCP~\cite{PhysRevLett.96.086805}.
As shown in Fig.~\ref{FIG4}\textbf{h}, upon applying an electric field, the current initially jumps to the geometric value $J = E_\mathrm{field}e^2 d_\mathrm{max}^2 / 8\hbar$, but continues to increase over time.
This extended rise originates from the finite group velocity induced by trigonal warping, resulting in a delayed response similar to monolayer graphene (Fig.~\ref{FIG1}\textbf{b}).
This effect can be suppressed by opening a gap at the linear BCPs.
Applying a perpendicular electric field $E_\mathrm{v}$ opens a tunable band gap (Fig.~\ref{FIG4}\textbf{d}), which eliminates the slowly growing intraband current and restores nearly instantaneous current generation driven by interband coupling (Fig.~\ref{FIG4}\textbf{h}).
As in V$_{3}$F$_{8}$, the rise time remains on the order of a few femtoseconds, while the steady-state current decreases with increasing $T_{2}$ or gap size. 
This ultrafast initial current surge enables current switching at petahertz-scale frequencies in BL graphene, as demonstrated in the following subsection.

\begin{figure*}[t]
\includegraphics[width=1.00\textwidth]{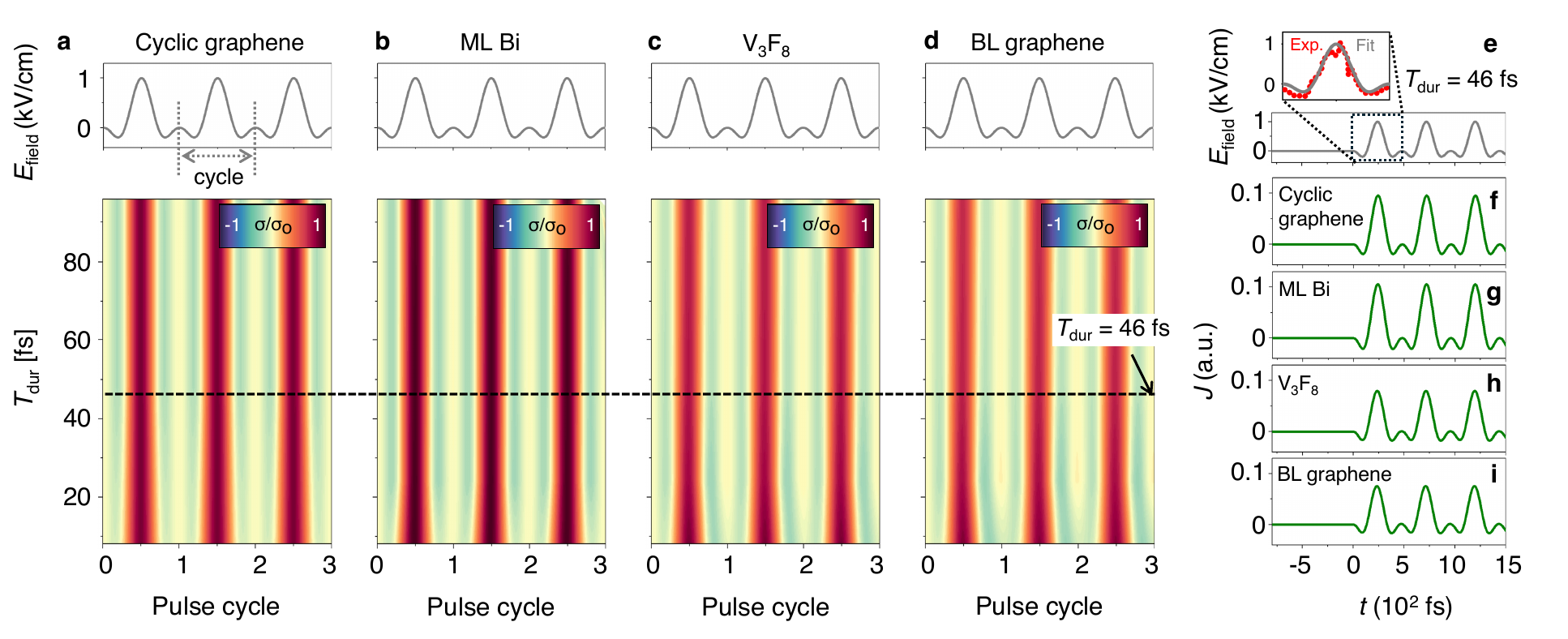}
\caption{
\textbf{Current switching behavior as a function of pulse duration.}
Current switching operation for (a) cyclic graphene, (b) ML Bi, (c) V$_{3}$F$_{8}$, and (d) BL graphene with 0.08 V/$\text{\AA}$ of $E_\mathrm{v}$ as a function of pulse duration $T_\mathrm{dur}$.
%
The systems are driven by a series of optical pulses with an electric field of 1 kV/cm, which is illustrated in the upper panel.
The color scale indicates the normalized conductivity, $\sigma(t)/\sigma_0$, showing the response during the "on" (red) and "off" (yellow) states.  
Here, $\sigma(t) = J(t)/E_\mathrm{field}$ and $\sigma_0 = e^2d_\mathrm{max}^2 /8\hbar$, with $d_{\mathrm{max}} = 1$.
(e) Optical pulse sequences and corresponding current responses for (f) cyclic graphene, (g) ML Bi, (h) V$_3$F$_8$, and (i) BL graphene. The currents are induced by an optical pulse sequence with $T_\mathrm{dur}=46$ fs, matching an experimentally observed pulse duration (red curve in an inset of (e))\cite{pulsefit2022}.
All simulations were performed with $T_{1}=100$ fs and $T_{2}=20$ fs.
}
\label{FIG5}
\end{figure*}

\subsection{Current Switching via Optical Pulse Trains}

Beyond the step-like electric field considered above, we assess the experimental feasibility of ultrafast current switching in the candidate materials under realistic optical pulses (Fig.~\ref{FIG5}).
By varying pulse durations ($T_\mathrm{dur}$) at a field intensity of 1\,kV/cm, Figs.~\ref{FIG5}\textbf{a-d} show the normalized conductivity $\sigma(t)/\sigma_0$ as a function of $T_\mathrm{dur}$, where $\sigma_0 = e^2 d_\mathrm{max}^2 / 8\hbar$ with $d_\mathrm{max}=1$.
The color scale distinguishes field-on (red) and field-off states (yellow), revealing robust current switching over a wide range of $T_\mathrm{dur}$ in both gapless (cyclic graphene and ML Bi) and gapped systems (V$_3$F$_8$ and BL graphene), with the current following the applied field instantaneously even as $T_\mathrm{dur}$ approaches several tens of femtoseconds.
Figure~\ref{FIG5}\textbf{e} shows a realistic optical pulse sequence with $T_\mathrm{dur}=46$\,fs, modeled by fitting experimentally observed pulses~\cite{pulsefit2022}, and the corresponding currents for all four systems are shown in Figs.~\ref{FIG5}\textbf{f-i}.
The switching behavior is universal, though gapless systems reach currents of $\sigma_0$ while gapped systems yield slightly reduced values, consistent with Fig.~\ref{FIG3}.
In all materials, the current faithfully follows the on-off profile of the optical field, confirming that the switching can be realized under experimentally relevant conditions, including pulse durations shorter than 46\,fs.
Recent experiments have demonstrated half-cycle sub-100\,fs pulses at kV/cm fields~\cite{pulsefit2022}, sub-picosecond pulse trains~\cite{tian_pulse}, and arbitrary THz pulse trains with tunable repetition rates~\cite{Abdo_pulse}, placing closely spaced half-cycle pulse bursts in the kV/cm regime well within experimental reach.

\section{Discussion}

In summary, we propose that QGSs with a BCP characterized by quantum geometry offer a compelling platform for realizing ultrafast electronic switching in next-generation devices.
QGSs exhibit instantaneous current generation via quantum geometric interband transitions, rapidly settling into a steady state upon application of a bias field.
This conclusion holds across a broad range of driving protocols, from square-wave bias fields to experimentally relevant optical pulse sequences, and remains valid at moderate electric field strengths, well below those required in all-optical light-wave control approaches~\cite{nature_rev_matt}.
This interband mechanism, fundamentally distinct from conventional intraband carrier dynamics, underlies the dramatic enhancement in switching speed achievable in QGSs.
Moreover, the steady-state conductivity is governed by the quantum distance, reflecting the underlying geometry of the Bloch states, and exhibits universal behavior independent of material-specific details.
These features remain robust even in the presence of a small energy gap.
Our findings are strongly supported by first-principles DFT calculations on realistic materials---cyclic graphene, ML Bi, V$_3$F$_8$, and AB-stacked bilayer graphene---all exhibiting current switching capabilities approaching the petahertz regime.

Among the various material candidates, BL graphene stands out as particularly promising given its extensive prior synthesis and characterization.
Its tunable band gap via a perpendicular electric field also provides an effective means to mitigate finite-temperature effects: thermal excitation partially populates the conduction band, enabling intraband transitions akin to conventional metals that degrade switching speed.
Opening a band gap suppresses this effect by reducing thermally excited conduction-band electrons, allowing interband processes to dominate.
In BL graphene, such gap control is intrinsically achievable via a perpendicular electric field, as discussed in the previous section.
As a result, BL graphene under $E_\mathrm{v} = 0.32$\,V/\text{\AA} exhibits robust current switching signals preserved up to 300\,K (see Supplementary Sec.~10).

To further validate our results within a more complete treatment where both occupation numbers and the electronic structure evolve self-consistently in real time, we perform ab initio real-time TDDFT calculations for ML Bi and cyclic graphene.
The current generation behavior of QGSs is consistently reproduced in these calculations (see Supplementary Sec.~12), reinforcing the robustness of our results and the viability of these systems as an electronics platform.
Note that since TDDFT does not include bath interactions, it cannot reproduce the current generation obtained for gapped QBT systems within the master equation approach.

Finally, we discuss robustness against disorder, a factor essential for device viability.
Under the Boltzmann transport framework, impurity scattering is elastic and modifies only the diagonal (intraband) populations in the density matrix, leaving the off-diagonal terms that generate interband current unaffected.
The interband current thus remains identical to the clean limit, while the intraband contribution is suppressed in QGS systems to leading order in the electric field (see Supplementary Sec.~11).
Consequently, the instantaneous current response and ultrafast switching behavior are expected to persist even in disordered QGS systems, confirming that QGSs provide a robust platform for ultrafast current switching under realistic conditions of temperature and disorder.

\section{Appendix}
\subsection{Quantum geometric semimetals and master equation}
We focus on QGSs characterized by $d_\mathrm{max}$ and exhibiting isotropic band dispersions.
Many QGSs, such as graphene bilayer, kagome lattice, and strained Bi monolayer, belong to this category.
Up to a unitary transformation, the generic form of the low-energy continuum Hamiltonian of such systems is given by
\begin{align}
\label{Eq1}
H_\mathbf{k}^\mathrm{QGS} = \big(f_x({\mathbf{k}}) + m_0\big)\sigma_x + f_y({\mathbf{k}}) \sigma_y + f_z({\mathbf{k}}) \sigma_z + (1-\eta)f_0({\mathbf{k}}) \sigma_0, 
\end{align}
where ${\sigma}_{x,y,z}$ represent the Pauli matrices, $\sigma_0$ denotes the identity operator.
Here, $f_x({\mathbf{k}})$, $f_y({\mathbf{k}})$, $f_z({\mathbf{k}})$, and $f_0({\mathbf{k}})$ are real-valued parabolic functions of $\mathbf{k}$ given by
\begin{align}
    f_x(\mathbf{k}) &= 2 {t_1} d_{\text{max}} \sqrt{1 - d_{\text{max}}^2}k_y^2, \\
    f_y(\mathbf{k}) &= 2 d_{\text{max}} {t_1} k_x k_y, \\
    f_z(\mathbf{k}) &= {t_1} \big[ k_x^2 +  (1 - 2 d_{\text{max}}^2) k_y^2\big], \\
    f_0(\mathbf{k}) &= {t_1} \big[ k_x^2 +  k_y^2 \big], 
\end{align}
where $t_1$ is the overall parameter describing mass tensors of quadratic bands and $m_0$ is a mass opening a band gap at the BCP.
The real parameter $\eta$ controls the effective masses of the upper and lower parabolic bands.
For example, when $\eta = 1$, the upper and lower quadratic bands possess identical effective masses but opposite curvatures.
In contrast, when $\eta = 0$, the lower band becomes completely flat while the upper band remains quadratic, yielding a singular flat band (SFB) system~\cite{Nature.584.59-63}.
We note that QBT systems with $0 \leq \eta \leq 2$ belong to the semimetals (see Supplementary Sec.~1).
Meanwhile, $d_\mathrm{max}$ solely determines the strength of interband coupling between two parabolic bands. 
From the eigenvectors obtained from $H_\mathbf{k}^\mathrm{QGS}$, one can confirm that the maximum value of the quantum distance is $d_\mathrm{max}$.
When $d_\mathrm{max} = 0$, the QGS becomes non-singular, leading to a complete suppression of interband coupling. 
Conversely, $d_\mathrm{max} = 1$ indicates maximum singularity, maximizing the interband coupling.
In total, there are four independent parameters, $\{t_1, \eta, d_\mathrm{max}, m_0\}$, which are consistent with the requirement of two effective masses for the isotropic conduction and valence bands, one interband coupling parameter, and one additional mass parameter.
Since any isotropic QBT system can be unitarily transformed into this form, the model captures all isotropic QGSs.
The model can be further generalized to anisotropic QGSs with distinct effective masses along the $x$- and $y$-axes, as detailed in Supplementary Sec.~1.

To describe the real-time dynamics, we employ the quantum master equation given by
\begin{align}
\frac{\partial}{\partial t}\rho_{\mathbf{k}}(t) = -\frac{i}{\hbar}\left[H_{\mathbf{k}-\frac{e}{\hbar c}\mathbf{A}(t)}, \rho_{\mathbf{k}}(t)\right] + \hat{D}[\rho_{\mathbf{k}}(t)],
\end{align}
where $H_{\mathbf{k}}$ is the Bloch Hamiltonian at crystal momentum $\mathbf{k}$ and $\rho_{\mathbf{k}}(t)$ is the time-dependent density matrix.
The system is coupled to an external electric field through the vector potential $\mathbf{A}(t) = c\mathbf{E}_{\text{field}}t\Theta(t)$, with $\mathbf{E}_{\text{field}} = \hat{x}E_{\text{field}}$.
Here, the step function $\Theta(t)$ is introduced to model the abrupt onset of the electric field.
For an optical pulse, the vector potential within the time window $-d_l \le t \le d_l$ is given by $ \mathbf{A}(t) = -\frac{1}{c}\int^{t} dt'\,\mathbf{E}_{\mathrm{field}}(t') $, where $ \mathbf{E}_{\mathrm{field}}(t) = E_{\mathrm{field}}\cos(\omega t)\cos^2\!\left(\frac{\pi t}{2d_l}\right) $, with $\omega = 0.05\,t_1$ and $d_l = 60/t_1$.
In this case, temporal duration of gaussian deviation can be defined as 

$T_{\mathrm{dur}}=\sqrt{\int dt t^2 \cos^4\left(\frac{\pi t}{2d_l}\right)\cos^2(\omega t)/\int dt \cos^4\left(\frac{\pi t}{2d_l}\right)\cos^2(\omega t)}$.

We also incorporate the coupling between the system and a bath via the dissipator $\hat{D}[\rho_{\mathbf{k}}(t)]$, where two relaxation times are introduced to describe nonequilibrium population decay ($T_1$) and dephasing ($T_2$), respectively.

The band structure of a simple metal is modeled by a quadratic dispersion corresponding to a typical doped semiconductor.
Graphene is described by linear bands crossing at the Dirac point.
Applying the same quantum master equation to these systems yields time-dependent current responses that smoothly approach steady-state values, with finite rise times governed by the population relaxation time $T_1$.
%


%
%

To incorporate a phenomenological description of bath, we use the Houston states, i.e., instantaneous eigen basis of $H_{\mathbf{k}-\frac{e}{\hbar c}\mathbf{A}(t)}$.
Therefore, the electron-bath coupling term within the instantaneous eigen basis can be written,
\begin{align}
\hat{D}_{H}[\rho_{\textbf{k}}(t)] = -\left( \begin{array}{cc}
\frac{\rho_{++,\textbf{k}}(t) - \rho_{+,\textbf{k}}^0(t)}{T_1} & \frac{\rho_{+-,\textbf{k}}(t)}{T_2} \\
\frac{\rho_{-+,\textbf{k}}(t)}{T_2} & \frac{\rho_{--,\textbf{k}}(t) - \rho_{-,\textbf{k}}^0(t)}{T_1}
\end{array} \right)
\end{align}
Here, $\hat{D}_{H}[\rho{\mathbf{k}}(t)]$ introduces the bath relaxation parameters $T_{1}$ and $T_{2}$, representing the population decay time and the dephasing time, respectively. The indices “$+$” and “$-$” refer to the top and bottom states. We assume that initially $\rho_{-,\textbf{k}}^0(t)=1$ (fully occupied) and $\rho_{+,\textbf{k}}^0(t)=0$ (empty).
To restore the coupling term to the original basis, we apply an inverse unitary transformation at each time step as $\hat{D}[\rho_{\mathbf{k}}(t)] = R_{\mathbf{k}(t)}\hat{D}_{H}[\rho_{\textbf{k}}(t)]R_{\mathbf{k}(t)}^{\dagger}$, where $R_{\mathbf{k}(t)}$ is the unitary matrix that diagonalizes $H_{\mathbf{k}-\frac{e}{\hbar c}\mathbf{A}(t)}$.
The current density is calculated by $\mathbf{J}(t) = (1/2\pi)^2\int d^{2}\mathbf{k}\left\langle {\mathbf{j}}_{\mathbf{k} - \frac{e}{\hbar c} \mathbf{A}(t)} \right\rangle$, where $\left\langle {\mathbf{j}}_{\mathbf{k} - \frac{e}{\hbar c} \mathbf{A}(t)} \right\rangle = \mathrm{Tr}[\rho_{\mathbf{k}}(t){\mathbf{j}}_{\mathbf{k} - \frac{e}{\hbar c}\mathbf{A}(t)}]$ and the current operator is given by ${\mathbf{j}}_{\mathbf{k} - \frac{e}{\hbar c}\mathbf{A}(t)} = -e\partial_{\mathbf{k}}H_{\mathbf{k}-\frac{e}{\hbar c}\mathbf{A}(t)}$.
In this study, we focus on the case $\mathbf{E}_{\text{field}} = {E}_{\text{field}}\hat{x}$, so the resulting current is along the $x$-direction: $\mathbf{J} = J\hat{x}$.

Additionally, calculations for other systems are performed using the respective Hamiltonians: for a simple metal, $H_\mathbf{k} = 2t_{1}\mathbf{k}^{2}$, and for graphene, $H_\mathbf{k} = {v_\mathrm{F}}(\chi{k}{x}\sigma_x+{k}{y}\sigma_y)$, where $\chi=1$ and $\chi=-1$ correspond to the Dirac cones at $\mathbf{K}$ and $\mathbf{K}^{\prime}$, respectively, and ${v_\mathrm{F}}=3\times10^{6}$ m/s~\cite{nphys2049}.
For the graphene case, the initial occupation ($\rho_{nn,\mathbf{k}}^0(t)$) assumes a fully filled lower cone and an empty upper cone.
On the other hand, for simple metal, a small carrier density is introduced via the Fermi-Dirac distribution, i.e.,  $\rho_{\mathbf{k}}^0(t)=f_\mathrm{FD}(\epsilon^{\mathbf{k}(t)}-\mu)$ with $\mu=1.3t_{1}$ and $\epsilon_{\mathbf{k}}=2t_{1}\mathbf{k}^{2}$. 
Especially for graphene, the total current includes contributions from both Dirac cones, i.e.,  $\mathbf{J} = \mathbf{J}(\chi=1) + \mathbf{J}(\chi=-1)$.
In the continuum model calculation, the bath term is assumed to be $T_{1} =$ 150$t_{1}^{-1}$ and $T_{2} =$ 30$t_{1}^{-1}$, unless otherwise specified.

\subsection{Master equation for the first-principles calculation}
To compute the time-dependent electron dynamics from first-principles under an external electric field, we employ the master equation in the length gauge, starting from the diagonalized field-free Hamiltonian, given by
\begin{align}
\frac{\partial}{\partial t}\rho_{{mn},\mathbf{k}}(t) 
&= -\frac{i}{\hbar}\left( \epsilon_m^{\mathbf{k}} - \epsilon_n^{\mathbf{k}} \right) \rho_{{mn},\mathbf{k}}(t) \nonumber \\
&\quad -\frac{i}{\hbar}\mathbf{E}_{\text{field}}(t) \cdot \sum_l \left[ \mathbf{d}_{ml}^{\mathbf{k}} \rho_{{ln},\mathbf{k}}(t) - \mathbf{d}_{ln}^{\mathbf{k}} \rho_{{ml},\mathbf{k}}(t) \right] \nonumber \\
&\quad + \frac{1}{\hbar}\mathbf{E}_{\text{field}}(t) \cdot \frac{\partial}{\partial \mathbf{k}} \rho_{{mn},\mathbf{k}}(t) \nonumber \\
&\quad + \hat{D}[\rho_{\mathbf{k}}(t)]_{mn}.
\end{align}
where $\epsilon^{\mathbf{k}}_{n} $ and $ \mathbf{d}^{\mathbf{k}}_{nm}$ stand for $n$-th eigenvalues and dipole matrix elements ($n$,$m$) at the momentum $\mathbf{k}$, respectively, directly calculated by the first-principles calculations.
For the multiband case, the bath term is given by
\begin{align}
\hat{D}[\rho_{\textbf{k}}(t)]_{mn} = -\delta_{mn}\frac{\rho_{mn,\textbf{k}}(t) - \rho_{n,\textbf{k}}^0}{T_1}-(1-\delta_{mn})\frac{\rho_{mn,\textbf{k}}(t)}{T_2}.
\end{align}
The initial occupations used in the main text are $\rho_{+,\textbf{k}}^0=0$ and $\rho_{-,\textbf{k}}^0=1$.
The bias electric field is given by $\mathbf{E}_{\text{field}}(t)=\mathbf{E}_{\text{field}}\Theta(t)$, thus the $\rho_{mn,\mathbf{k}}(t=0)=\rho_{m,\mathbf{k}}^{0}\delta_{mn}$.
The current density is computed as $\mathbf{J} = (1/2\pi)^2\int d^{2}\mathbf{k}\left\langle {\mathbf{j}}_{\mathbf{k}} \right\rangle$.
All other procedures follow the same treatment as in the master equation for the continuum model.

\subsection{First-principles calculations}
We perform DFT calculations by using \textsc{Quantum Espresso}~\cite{JPhys_CM_29_465901_2017} with the plane-wave (PW) basis, the PBE exchange-correlation functional~\cite{PhysRevLett_77_3865_1996} and norm-conserving pseudopotentials~\cite{ComPhysComms_226_39_2018_Setten}. 
Self-consistency is achieved by using $27\times27\times1$ $k$-point mesh, and the kinetic energy cutoff $400$ Ry. 
We insert a vacuum space 15 $\textrm{\AA}$ to prevent interactions between periodic image layers. 

The lattice constants for cyclic graphene, ML Bi, V$_3$F$_8$, and BL graphene are taken to be 5.19 $\textrm{\AA}$, 4.63 $\textrm{\AA}$, 6.34 $\textrm{\AA}$, and 2.47 $\textrm{\AA}$, respectively. 
For the ML Bi, the fully relaxed structure with a lattice constant of 4.40 $\textrm{\AA}$ exhibits a finite band gap at the $\Gamma$ point. 
To close the band gap, a biaxial strain of approximately 5.2 $\%$ was applied, corresponding to a lattice constant of 4.63 $\textrm{\AA}$.

For the magnetic calculation of V$_{3}$F$_{8}$ we adopt the DFT+$U$+$V$ method where on-site ($U$) and inter-site ($V$) Hubbard interactions are self-consistenly determined in combination with DFT calculations~\cite{PhysRevResearch.2.043410}. 
For ML Bi, spin-orbit coupling is included in the electronic 
structure calculations, while for V$_{3}$F$_{8}$, spin polarization 
is also taken into account.
After the DFT calculations, $\epsilon^{\mathbf{k}}_{n} $ and $ \mathbf{d}^{\mathbf{k}}_{nm}$ are plugged into the master equation through the wannierization of $300\times300$ sampled first Brillouin zone.
The Wannier functions for each system are constructed as follows.
For cyclic graphene, 15 Wannier basis functions are employed, 
comprising $p_z$ orbitals at each C atom and six $\sigma$-bonding 
($sp^2$-derived) orbitals between C atoms.
For ML Bi, 12 Wannier basis functions are used, 
comprising $p$ orbitals at each Bi atom.
For V$_3$F$_8$, 15 Wannier basis functions are employed, 
comprising $p$ and $d$ orbitals for V atoms and $p$ orbitals for F atoms.
For BL graphene, 10 Wannier basis functions are employed, 
comprising $p_z$ orbitals at each C atom and six $sp^2$-derived $\sigma$-bonding orbitals between C atoms.

\subsection{The expression of $\mathcal{G}(\mathbf{k},d_\mathrm{max},m_{0}T_{2})$ in  current density}

With $q_{x} \equiv \sqrt{\frac{t_1}{m_0}}k_{x}$ and $q_{y} \equiv \sqrt{\frac{t_1}{m_0}}k_{y}$, the explicit form of $\mathcal{G}(\mathbf{k},d_\mathrm{max},m_{0}T_{2})$ function in the current density is given by, 
\begin{eqnarray}
    \mathcal{G}(\mathbf{k},d_\mathrm{max},m_{0}T_{2}) &=&  \frac{\left(\left[bq_{y}g_{y}g_{z}-2q_{x}\left(g_{x}^{2}+g_{y}^{2}\right)\right]^{2}+\left[bq_{y}g_{x}|\mathbf{g}|\right]^{2}\right)}{\Big(|\mathbf{g}|^{3}\left(g_{x}^{2}+g_{y}^{2}\right)\Big)\Big(|\mathbf{g}|^{2}+\left(\frac{1}{2m_{0}T_{2}}\right)^{2}\Big)}. \nonumber\\
\end{eqnarray}
Here, $g_{x} = \left(aq_{y}^{2}+1\right)$, $g_{y} = b q_{x} q_{y}$, $g_{z} = q_{x}^{2} + c q_{y}^{2}$, $ \left|\mathbf{g}\right| = \sqrt{g_{x}^{2} + g_{y}^{2} + g_{z}^{2}}$, $a=2d_{\text{max}}\sqrt{1-d_{\text{max}}^{2}}$, $b=2d_{\text{max}}$, and $c=1-2d_{\text{max}}^{2}$.
Further details of the derivation can be found in Supplementary Sec. 5.

\subsection{The rise time formula}
The finite rise time can be roughly estimated using the uncertainty principle, $\Delta t \Delta E \geq \hbar$.
Taking $\Delta E$ to be the energy scale associated with the bandwidth of the parabolic portion of the band structure—of the order of $\sim$eV—we find that the lowest bound of $\Delta t$ is of the order of a few femtoseconds, consistent with the simulations.
Beyond this rough estimation, we derive an analytic expression for the rise time in the presence of a finite momentum cutoff in the current density integration, given by $W\left(\frac{10\hbar}{\pi t_1 T_2 k_c^2}\right) T_2$, where $W(x)$ is the Lambert W-function (see Supplementary Sec. 6A).
Unlike the simple uncertainty-based estimate, the analytic expression reveals a strong dependence on the dephasing time $T_2$; however, for typical values of $T_2$, the rise time remains on the order of a few femtoseconds.
Despite the differences in early-time dynamics between the continuum and DFT results, the steady-state current density is identical ($J = E_\mathrm{field}e^2d_\mathrm{max}^2 /8\hbar$) in both cases.

\section{Acknowledgement}
Y.K. was supported by a KIAS Individual Grant (PG088602) at Korea Institute for Advanced Study (KIAS) and supported by the National Research Foundation of Korea (NRF) grant funded by the Korea government (MSIT) (Grant no. RS-2025-00553820).
J.W.R was supported by the National Research Foundation of Korea (NRF) Grant funded by the Korean government (MSIT) (Grant nos. 2021R1A2C1010572 and 2022M3H3A1063074) and the Ministry of Education (Grant no. RS-2023-00285390).
S.K was supported by the National Research Foundation (NRF) of Korea (Grant no. RS-2024-00410027 and RS-2025-16065427).
S.K and J.W.R acknowledge the support from Korea Institute for Advanced Study (KIAS) grant funded by the Korea government.
Authors thank the computational support from the Center for Advanced Computation (CAC) at Korea Institute for Advanced Study (KIAS).

\textbf{Author contributions}
Y.K. conceived the original idea. Y.K., S.K, and J.-W.R. materialized the research idea, performed the theoretical calculation, discussed the results, and wrote the manuscript.

\textbf{Supplemental Material}

\bibliography{bib}

\clearpage

\clearpage

\clearpage
\clearpage

\clearpage
\clearpage

\clearpage

\end{document}